\documentclass[pdflatex,12pt]{article}

\usepackage{graphicx}
\usepackage{multirow}
\usepackage{amsmath,amssymb,amsfonts}

\usepackage[left=2.5cm, right=2.5cm, bottom=2cm]{geometry}
 \usepackage[margin=10mm,font=small,labelfont=bf]{caption}

 \newcommand{\sgn}{\operatorname{sgn}}

\begin{document}

\vspace{-32mm}
\hspace{3mm}\noindent
{\Large{\bf Exact solution of the Heat Equation\\ 

\vspace{-6mm}
\hspace{8mm}for initial polynomials or splines}}

\vspace{6mm}
\noindent 
\hspace{10mm}Mark Andrews

\vspace{2mm}
\hspace{4mm}\noindent 
Department of Quantum Science, Research School of Physics\\

\vspace{-5mm}
\hspace{4mm}\noindent 
The Australian National University, Canberra  ACT 2600, Australia\\

\vspace{-4mm}\noindent
\hspace{11mm}email: Mark.Andrews@anu.edu.au

\vspace{3mm}
\abstract{The exact evolution in time and  space of a distribution of the temperature (or density of diffusing matter) in an isotropic homogeneous medium is determined where the initial distribution is described by a piecewise polynomial. In two dimensions, the boundaries of each polynomial must lie on a grid of lines parallel to the axes, while in three dimensions the boundaries must lie on planes perpendicular to the axes. The distribution at any position and later time is expressed as a finite linear combination of Gaussians and Error Functions. The underlying theory is developed in detail for one, two, and three dimensional space, and illustrative examples are examined.}


\section{Introduction}\label{sec1}

The heat equation 
\vspace{-2mm}
\begin{equation}\label{eqn:Heat}
\partial_t \phi =\kappa\, \nabla^2 \phi,
\end{equation}
describes how a distribution of temperature $\phi$ varies with space and time in an isotropic homogeneous medium. The same equation applies to the diffusion of a distribution of matter with density $\phi$. The mathematical solution of the heat equation can be found directly in simple cases and the fundamental solution (or propagator)\cite{F}
\begin{equation}\label{eqn:chi}
\hspace{1cm}\chi(x,t)=\frac{1}{\sqrt{4\pi \kappa t}}\exp(-\frac{\,x^2}{4\kappa t})
\end{equation}
gives a formal solution as an integral over the initial state:
\begin{equation}\label{eqn:ev}
\hspace{2cm}\phi(x,t)=\int \chi(x-x',t)\phi(x',0) \textrm{d}x',
\end{equation}
and this extends to more dimensions where the propagator is the product of the one-dimensional form for each Cartesian dimension. Since this integral can be carried out exactly only in special cases, iterative numerical methods are frequently applied. Here we show that exact solutions can be found when the initial state is described by a piecewise polynomial. Such polynomials can be used to describe any finite initial state to arbitrary accuracy.\cite{SW} It will be shown that their evolution can be expressed as a finite combination of Gaussians and Error Functions.

\pagebreak
\section{The one-dimensional case}\label{sec2}
Our piecewise polynomials of degree $n$ consist of $N$ segments, each described by a polynomial in $x$ of degree less than or equal to $n$. The method involves integrating (\ref{eqn:ev}) by parts repeatedly, thereby introducing the sequence of antiderivatives of the propagator $\chi(x,t)$. This sequence of repeated integrations by parts ends because the $(n+1)$th derivative of the polynomial must be zero. This results in an expression for the time-evolved state as a finite combination of the repeated integrals of the propagator evaluated at the boundaries of the segments.

\subsection{Solution of the Heat Equation in terms of the integrals of the propagator}
The initial state $\phi(x,0)$ is assumed to be zero outside the range $a_0\leq x \leq a_N$ and for segment $i$ with $a_{i-1}<x<a_{i}$, $\phi(x,0)=u_{i}(x)$, a polynomial of degree $\leq n$. At least one of $u_i(x)$ and its derivatives will differ from each of its immediate neighbours at the boundaries. Smoothness conditions, such as continuity of  $\phi(x,0)$ and some of its derivatives, might be imposed; but that is not required here.

Taking $u_i(x)$ to be zero outside $a_{i-1}\!<x<a_{i}$, the evolution equation (\ref{eqn:ev}) becomes
\vspace{-3mm}
\begin{equation}
\phi(x,t)=\sum_{i=1}^{\;N}\int_{a_{i-1}}^{a_i}\chi(x-x')u_i(x')\,\textrm{d}x'.
\end{equation}

\vspace{-3mm}
\noindent We now introduce the sequence of antiderivatives of the propagator,
\begin{equation}\label{eqn:chip}
\chi_{p+1}(x,t):=\int_0^x \chi_p(x',t)\textrm{d}x'
\end{equation}

\vspace{-3mm}
\noindent starting with $\chi_{0}(x,t):=\int_0^x \chi(x',t)\textrm{d}x'$. Denoting the $p$th derivative of $u_i(x)$ by $u^p_i(x)$, integration by parts gives
\begin{eqnarray}\nonumber
\hspace{-10mm}\int_{a_{i-1}}^{a_i}\!\chi_p(x-x') u^{p+1}_i(x')\,\textrm{d}x'\!=\!\int_{a_{i-1}}^{a_i}\!\chi_{p+1}(x-x')u^p_i(x')\,\textrm{d}x'+\label{eqn:IntByParts}\\
\hspace{35mm}\chi_{p+1}(x-a_{i-1})u^{p+1}_i(a_{i-1})-\chi_{p+1}(x-a_{i})u^{p+1}_i(a_{i}).
\end{eqnarray}
Starting with $p=-1$ and $\chi_{-1}(x,t)\equiv\chi(x,t)$, this operation can be repeated until the integrand becomes zero for some $p\leq n$, (noting that $u_i^{n+1}(x)\equiv0$). Therefore
\begin{equation}\label{eqn:phi_chi1}
\phi(x,t)=\sum_{i=1}^N\sum_{p=0}^n C_i^p\chi_p(x,t),
\end{equation}
where $C_i^p=\lim_{\epsilon\to 0} [\phi^p(a_i+\epsilon)-\phi^p(a_i-\epsilon)]$. That is, $C_i^p$ is the jump in the $p$th derivative of the initial state $\phi(x)$ at the point $x=a_i$; it will be non-zero only at discontinuities in this derivative. [The symbols $p, q, r$ are used to denote indeces, never powers.]

\pagebreak
\subsection{Repeated integrals of the propagator}
We need the explicit form of the sequence of functions, $\chi_{p+1}(x,t)=\int_0^x\,\chi_p(y,t)\,\textrm{d}y$ with
 $\chi_{-1}(x,t)\equiv\chi(x,t)$. The Error Function, is defined to be
\begin{equation}\label{eqn:E0}
\mathcal{E}_0(z)\equiv\,\textrm{erf}(z):= \frac{2}{\sqrt{\pi}}\int_0^z\!\exp(-y^2)\,dy,
\end{equation}
and hence $\chi_0(x,t)=\frac{1}{2}\,\mathcal{E}_0(x/\sqrt{4\kappa t}).$
With $\mathcal{E}_{-1}(z):=2\pi^{-1/2}\exp(-z^2)$ and $\mathcal{E}_0(z)=\textrm{erf}(z)$, we continue the sequence 
\begin{equation}
\mathcal{E}_p(z):=\int_0^z \mathcal{E}_{p-1}(z')\,\textrm{d}z',\;\;\;\;\textrm{d}_z\mathcal{E}_p(z)=\mathcal{E}_{p-1}(z)
\end{equation}
and it is easy to check that
\begin{equation}\label{eqn:Eseq}
p\,\mathcal{E}_p(z)=[z\,\mathcal{E}_{p-1}(z)+\mathcal{E}_{p-2}(z)]\;\;\;\textrm{for}\;p\geq1,
\end{equation}
and $\mathcal{E}_p(z)$ is antisymmetric for even $p$, symmetric for odd $p$, and $\mathcal{E}_p(0)=0$ for $p\geq 0$.

\noindent[The sequence in (\ref{eqn:Eseq}) is similar to that for the repeated integrals of $1-\textrm{erf}(z)$.\cite{M}]

It follows that the required sequence of integrals of the propagator is $\chi_p(x,t)$ with
\begin{equation}
p\,\chi_p(x,t)=x\,\chi_{p-1}(x,t)+2\kappa\,t\,\chi_{p-2}(x,t)
\end{equation}
with the property that $\partial_x\chi_p(x,t)=\chi_{p-1}(x,t)$ for $p\geq  0$ and  $\chi_{-1}(x,t)\equiv\chi(x,t)$. 

Every $\chi_p(x,t)$ can therefore be expressed as a finite combination of the Gaussian $\exp(-x^2/4\kappa t)$ and the Error Function $\textrm{erf}(x/\sqrt{4\kappa t})$; no other transcendental functions are required to evolve any initial piecewise polynomial.

The asymptotic form of the Error Function is $\textrm{erf}(z) \to \pm 1\; \textrm{as}\; z \to \pm\infty$; therefore 
\begin{equation}\label{eqn:fp}
\hspace{-1cm}\chi_p(x,t)\;\to\;f_p(x):=\frac{1}{2p!}x^{p-1}|x|\;\;\textrm{as}\;t\,\to\,0\;\textrm{for }p\geq 0\; \textrm{and any fixed } x.
\end{equation}

\subsection{Asymptotic form of the evolution as time increases}
For any bounded initial state that is non-zero only for a finite region, it is known \cite{V} that the solution tends to a Gaussian form. In some examples, we include for comparison a Gaussian with the same `mass' $M=\int\!\phi(x)\,\textrm{d}x$ and `m-width' $w$ (the mean deviation from the median) as that of the initial state $\phi(x)$. If the point with $x=m_1$ divides the first mass-quartile from the second (so that $\int_{-\infty}^{\,m_1}\phi(x)\textrm{d}x=\frac{1}{4}M$) and $x=m_3$ divides the third mass-quartile from the fourth, then the m-width of $\phi(x)$ is $w=\frac{1}{2}(x_3-x_1)$. The Gaussian $G(x)\!=\!M\exp(-x^2/\alpha^2)/(\alpha\sqrt{\pi})$ has $w\!\approx 0.476936\,\alpha$ (because erf$(0.476936) \approx 0.5$) and mass $M$. For an initial state with m-width $w$ the Gaussian with the same m-width has $\alpha=w/0.476936$. The Gaussian is to be centred at the median $\bar{x}=m_2$, such that 
$\int_{-\infty}^{\,m_2}\phi(x)\textrm{d}x=\frac{1}{2}M$. The exact evolution of  $G(x)$ is 
\begin{equation}
G(x,t)=\frac{M}{\sqrt{\pi(\alpha^2+4\kappa\,t)}}\exp\!\big(\!-\frac{x^2}{\alpha^2+4\kappa\,t}\big).
\end{equation}

\subsection{Alternative approach that will be used for two and three dimensions}
Instead of determining the evolution directly, it is possible to first express the polynomials as a linear combination of the functions $f_p(x)=\frac{1}{2}\sgn(x)x^p/p!$ in (\ref{eqn:fp}) that are the limit as $t\to 0$ of the integrals $\chi_p(x,t)$ of the propagator.  Then the evolution is obtained by replacing each $f_p(x)$ by $\chi_p(x,t)$, because each $\chi_p(x,t)$ also satisfies the heat equation. [If $(\partial_t -\kappa\, \nabla^2)\chi_p(x,t)=0$ then 
$(\partial_t -\kappa\, \nabla^2)\int_0^x\chi_p(x',t)\textrm{d}x'=0$.]

Applying Taylor's theorem to any polynomial $u(x)$ of degree $n$ leads to the result $\sum_{p=0}^n f_p(x-a)u^p(a)=\frac{1}{2}\sgn(x-a)u(x)$ for all $x$. Therefore, for any $\bar{a}>a$,
\begin{equation}
\sum_{p=0}^n [f_p(x-a)\,u^p(a)-f_p(x-\bar{a})\,u^p(\bar{a})]=\begin{cases}u(x),&\textrm{for}\;a< x< \bar{a}\\
0,&\textrm{otherwise},\end{cases}
\end{equation}
because the two terms add to give $u(x)$ for $a< x< \bar{a}$ and cancel otherwise. Thus if the initial state is described for all $x$ by the polynomial $u(x)$ truncated to the finite region $a< x< \bar{a}$, it can be replaced by an equivalent superposition of the global functions $f_p(x)$ and therefore will evolve to 
$\sum_{p=0}^n [\chi_p(x-a,t)\,u^p(a)-\chi_p(x-\bar{a},t)\,u^p(\bar{a})]$ for all $x$ and $t\geq 0$. 

If the initial state includes more such finite regions, each described by its own truncated polynomial, then the evolutions of each region can be added because the heat equation is linear. For a piecewise polynomial $\phi(x)$ such that it and possibly some of its derivatives are continuous, then it is best to express the evolution as in (\ref{eqn:phi_chi1}), because $C_i^p=0$ if $\phi^p(x)$ is continuous at $x=a_i$.

\vspace{8mm}
\subsection{Examples in one dimension}
All that we need to evolve a given piecewise polynomial $u(x)$ is the location and amount of the discontinuities in  $u(x)$ or any of its derivatives  $u_i^p(x)$. This information is contained in the list of points of discontinuity $a_i$ and the jump $C_i^p$ in $u^p(x)$ at $x=a_i$. For a continuous distribution, all $C_i^0=0$. If the distribution is smooth (no sudden change in slope at $a_i$) all $C_i^1=0$; but allowing a slope only at the end points may be required. A piecewise polynomial subject to such continuity conditions is called a spline; we deal with an example (a B-spline) below.

\pagebreak
\subsubsection{Initial square distribution}$u(x)=\frac{1}{2}$ for $|x|<a$ and zero otherwise; $a_0=-a$ with $C_0^0=\frac{1}{2}$ and $a_1=a$ with $C_1^0=-\frac{1}{2}$. Thus, the initial state can be written as $u(x)=\frac{1}{2}[f_0(x-a_0)- f_0(x-a_1)]$ and its evolution is $\phi(x,t)=\frac{1}{2}[\chi_0(x-a_0,t)- \chi_0(x-a_1,t)]$. The mass $M=1$, the median $\bar{x}=0$ and the m-width $w=\frac{1}{2}a$. This evolution is shown in Fig.(1a). 
Note that $\phi(x,t)$ does become close to the corresponding $G(x,t)$ for $t > a^2/\kappa$.

\subsubsection{Initial triangular distribution:} $u(x)=1-|x|/a$ for $|x|<a$ and zero otherwise; $a_0=-a$ with $C_0^1=1$, $a_1=0$ with $C_1^1=-2$, and $a_2=a$ with $C_2^1=1$. Thus, the initial state can be written as $u(x)=f_1(x-a_0)- 2f_1(x-a_1)+f_1(x-a_2)$ and its evolution is $\phi(x,t)=\frac{1}{2}[\chi_1(x-a_0,t)- 2\chi_1(x-a_1,t)+\chi_1(x-a_2,t)$. The mass $M=1$, the median $\bar{x}=0$ and the m-width $w=(1-\sqrt{2}/2)a$. This evolution is shown in Fig.(1b). This initial state is much more like a Gaussian than is the square state, and approaches the Gaussian asymptotic form much more rapidly.

\begin{figure}[h!]
\begin{centering}
\includegraphics[width=0.95\textwidth]{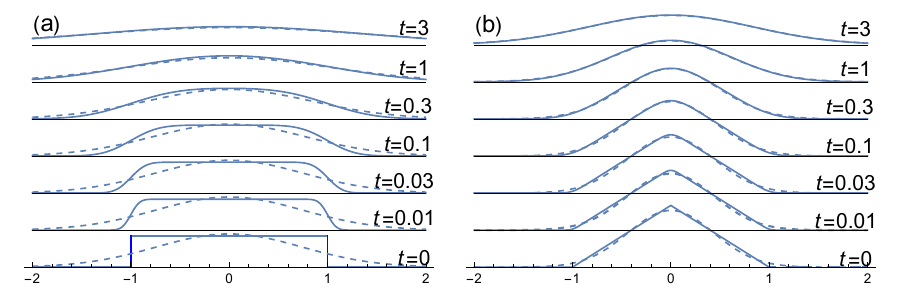}
\caption[left]{The evolution of two simple piecewise polynomials with the same mass: (a) a square initial distribution with a constant value for $|x|<a$, (b) a triangular distribution with density $1-|x/a|$. The results (solid curves) are shown for a sequence of times, together with the Gaussian with the same initial median mass and mean deviation of mass from the median, $w$. The units of length and time are $a$ and $a^2/\kappa$.}
\end{centering}  
\label{fig:1}
\end{figure}

\pagebreak
\subsubsection{Initial B-spline:} In the original cubic B-spline method\,\cite{K}, a data set of values $u_i$ with $i=1$ to $\scriptstyle{N}$, equally spaced over the range $[a,b]$ is interpolated by the function

\vspace{-3mm}
\begin{equation}\label{eqn:Balpha}
u(x)= \sum_{i=-1}^{\scriptstyle{N}+1}\alpha_i B_3\big(\frac{x-x_i}{d}\big)\;\;\;\textrm{for}\;x\in[a,b],
\end{equation}

\vspace{-1mm}\noindent
where $d=(b-a)/\scriptstyle{N}$ and $B_3(x)$ is the piecewise polynomial $\frac{1}{6}\big((2-|x|)^3-(4-|x|)^3\big)$ if $|x|\leq 1$, $\frac{1}{6}\big((2-|x|)^3\big)$ if $1\leq |x| \leq 2$, and $0$ if $|x|\geq 2$. To have $u(x)$ pass through the given data points and have the specified slopes at the end points, we require

\vspace{-3mm}
\begin{equation}\label{eqn:alpha}
\alpha_{i}\!+\!4\alpha_{i-1}\!+\!\alpha_{i+1}=6u_i,\;\;\;\;\alpha_1\!-\!\alpha_{-1}=2d\,u^1_0,
 \;\;\;\;\alpha_{\scriptstyle{N}+1}\!-\!\alpha_{\scriptstyle{N}-1}=2d\,u^1_{\scriptstyle{N}},
\end{equation}

\vspace{-1mm}\noindent
where $u^1_1=\partial_x u(x)_{x\to a}$ and $u^1_{\scriptstyle{N}}=\partial_x u(x)_{x\to b}$. The shape of $B_3(x)$ is shown in Fig.(2a).

\vspace{2mm}
Since $B_3(x)$ is a piecewise polynomial and only its third derivative is discontinuous, it is simple to express $u(x)$ in terms of the functions $f_3(x)$ and evolve it. The change in the third derivative of $B_3(x)$ at $x=\{-2,\,-1,\,0,\,1,\,2\}$ is $\{1, -4, 6, -4, 1\}$ and therefore $C^3_i=\alpha_{i-2}-4\alpha_{i-1}+6\alpha_{i}-4\alpha_{i+1}+\alpha_{i+2}$. This should be modified at the ends because the interpolation extends (to a small but significant extent) outside the range $\{a,b\}$. To avoid this error, the interpolation must be truncated at $a$ and $b$ by including its derivatives at the boundaries using $C^1_0=u^1_0,\;
C^2_0=\alpha_{-1}-2\alpha_{0}+\alpha_{1},\; C^3_0=\alpha_{-1}-3\alpha_{0}+3\alpha_{1}-\alpha_{2}$ and $C^1_{\scriptstyle{N}}=-u^1_{\scriptstyle{N}},\; 
C^2_{\scriptstyle{N}}=-(\alpha_{\scriptstyle{N}-1}-2\alpha_{\scriptstyle{N}}+\alpha_{\scriptstyle{N}+1}),\; 
C^3_{\scriptstyle{N}}=-(\alpha_{\scriptstyle{N}-2}-3\alpha_{\scriptstyle{N}-1}+3\alpha_{\scriptstyle{N}}-\alpha_{\scriptstyle{N}+1})$. Then the exact evolution of the truncated interpolation is (assuming $u(a)=u(b)=0$)
\vspace{-3mm}
\begin{equation}\label{eqn:Bevolve}
\phi(x,t)=\sum_{i=1}^{\scriptstyle{N}-1}C^3_i\,\chi_3(x-x_i)+\sum_{p=1}^{3}C^p_0\,\chi_p(x-a)+
\sum_{p=1}^{3}C^p_{\scriptstyle{N}}\,\chi_p(x-b).
\end{equation}

\vspace{-3mm}
For example, with $\scriptstyle{N}=10$, $u_i=\sin\!\big(\pi (i/10)^2\big)$ and $u^1_{\scriptstyle{N}}=0,\,u^1_{\scriptstyle{N}}=-\pi/5$, then solving (\ref{eqn:alpha}) for $\alpha_i$ leads to the B-spline interpolation in (\ref{eqn:Balpha}) and its evolution (\ref{eqn:Bevolve}) shown in Fig.2(b). For the Gaussian asymptotic form, we take mass $M=\int_0^{10} \sin\!\big(\pi (x/10)^2\big)\,\textrm{d}x\approx 5.05$, mass-median $\bar{x}=6.49$, and m-width $w=1.38$.

\begin{figure}[h!]
\begin{centering}
\includegraphics[width=0.93\textwidth]{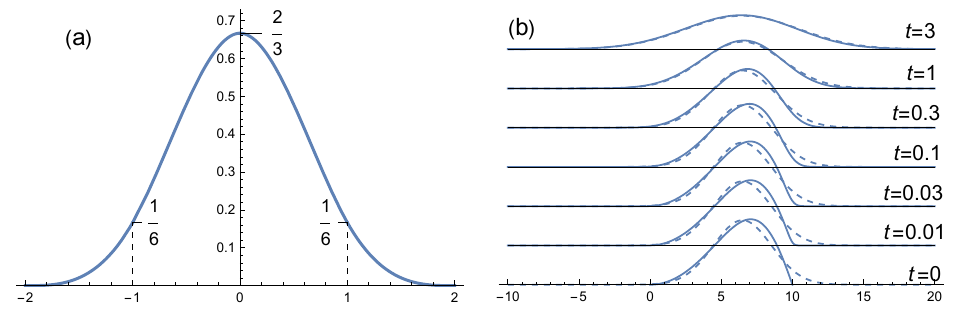} 
\caption[left]{(a) The $B_3$ basis spline. (b) The evolution of a B-spline interpolation. The results (solid curves) are shown for a sequence of times, together with the Gaussian with the same initial median mass and m-width $w$. The unit of time is $w^2/\kappa$.}
\end{centering}  
\label{fig:2}
\end{figure}

\pagebreak

\section{Two-dimensional media}
\vspace{0mm}
In one dimension, any initial piecewise polynomial can be evolved; but in a medium with two dimensions the previous method extends only to a set of polynomials each truncated to a rectangular region. For the efficiencies arising from any continuities across the boundaries to be realised, the boundaries for each polynomial must lie on a rectangular grid; but the spacings between the lines of the grid may vary. 

\subsection{Evolving a polynomial truncated by a rectangle}

\vspace{-1mm}
For an initial state $\phi(x,y,0)$ the evolution $\phi(x,y,t)$ can be found from the integral

\vspace{-3mm}
\begin{equation}\label{eq:ev2D}
\phi(x,y,t)=\mbox{$\int\!\!\int$}\,\chi(x-x',t)\,\chi(y-y',t)\, \phi(x',y',0) \,\textrm{d}x'\,\textrm{d}y'.
\end{equation}

\noindent
If the initial state $\phi(x,y,0)$ is a polynomial $u(x,y)$ of degree $n$ truncated to the rectangle $R=\{a_-\leq x\leq a_+,\;b_-\leq y\leq b_+\}$, the integral can be converted to a sum using Taylor's theorem: $\sum_{p=0}^n\sum_{q=0}^n f_p(x-a_\alpha)f_q(y-b_\beta) u^{p,q}_{\alpha,\beta} =\frac{1}{4}\sgn(x-a_\alpha)\sgn(y-b_\beta) u(x,y)$ for all $x, y$ where $u^{p,q}_{\alpha,\beta}:=u^{p,q}(a_\alpha,b_\beta)$ and $\alpha$ and $\beta$ can be + or $-$ to indicate which corner of the rectangle is used as the origin of the Taylor series. Therefore
\vspace{-2mm}
\begin{equation}\label{eqn:ev2Dff}
\sum_{p=0}^n \sum_{q=0}^n \sum_{\alpha=-}^+ \sum_{\beta=-}^+\sigma_{\alpha,\beta}
 \,f_p(x-a_\alpha)\,f_q(y-b_\beta)\,u^{p,q}_{\alpha,\beta}=
 \begin{cases}u(x,y),&\textrm{for}\;\{x,y\}\in R \\
0,&\textrm{otherwise},\end{cases}
\end{equation}

\vspace{-2mm}
\noindent where $\sigma_{\alpha,\beta}=\alpha\beta$, showing that the terms from the $(-,-)$-corner (with the lowest values of $x$ and $y$) are added, as are those from the (+,+)-corner, while the other two corners are subtracted. The evolution of the truncated polynomial is then
\vspace{-2.5mm}
\begin{equation}\label{eqn:ev2Dchi}
\phi(x,y,t)=\sum_{p=0}^n \sum_{q=0}^n \sum_{\alpha=-}^+ \sum_{\beta=-}^+\sigma_{\alpha,\beta}
 \,\chi_p(x-a_\alpha,t)\,\chi_q(y-b_\beta,t)\,u^{p,q}_{\alpha,\beta}.
\end{equation}

\vspace{-5mm}

\subsection{The evolution of piecewise polynomials truncated by a rectangular grid}
\vspace{-2mm}
For multiple polynomials, the simplest case has all the rectangular regions part of a grid made up of lines with $x=\{a_0, a_1, ..., a_{Nx}\}$ and lines with $y=\{b_0, b_1, ..., b_{Ny}\}$. The spacing  of the $x$-lines may vary and may differ from those of the $y$-lines. [More complex arrangements are possible, such as rectangular subregions with their own grid.]

Assume that $\phi(x,y,0)$ is defined within each rectangular segment $S_{i,j}$ with \mbox{$a_{i-1}<x<a_i$} and $b_{j-1}<y<b_j$ by a polynomial $u_{i,j}(x,y)$. For any of the segments, within the outer rectangle $a_0<x< a_{Nx},\,b_0<y< b_{Ny}$,  $u(x,y)$ may be zero throughout that segment. With $n$ denoting the largest degree of all the polynomials, the result is
\vspace{-2mm}
\begin{equation}\label{eqn:phi_chi2}
\phi(x,y,t)=\sum_{i=1}^{Nx} \sum_{j=1}^{Ny} \sum_{p=0}^n \sum_{q=0}^n C_{i,j}^{p,q}\,\chi_p(x-a_i,t)\,\chi_q(y-b_j,t),
\end{equation}

\vspace{-3mm}
\noindent where 
\begin{equation}
C_{i,j}^{p,q}=\sum_{\alpha=-}^+ \sum_{\beta=-}^+\sigma_{\alpha,\beta}\,
u_{\alpha,\beta}^{\;p,\,q}.
\end{equation}
Here $u_{\alpha,\beta}^{\,p,\,q}$ refers to the nearest corner of the four polynomials surrounding the point\,$(i,j)$. [$u_{i,j}$ and $u_{i+1,j+1}$ have $\sigma_{\alpha,\beta}=+1$ and $u_{i+1,j}$ and $u_{i,j+1}$ have $\sigma_{\alpha,\beta}=-1$.]
The subscripts $i,j$ in $u_{i,j}$ refer to the region $S_{i,j}$ while $C_{i,j}$ relates to the junction at the point $(a_i,\,b_j)$.

Continuity conditions ensure that many of the $C_{i,j}^{p,q}$ will be zero for splines. Thus, if 
$\phi(x,y,0)$ is continuous over the whole spline, then $p=0$ and $q=0$ are excluded because $f_0(x)$ jumps at $x=0$; from (\ref{eqn:phi_chi2}), $\phi(x,y,0)$ has jumps at $x=a_i$ for most values of $y$ if $C_{i,j}^{0,q}\neq 0$ and similarly for  $C_{i,j}^{p,0}$. Also, if 
$\partial_x\phi(x,y,0)$ is continuous over the whole spline, then $p=1$ is excluded because $\partial_xf_1(x)$ jumps at $x=0$ and hence $\partial_x\phi(x,y,0)$ has jumps at $x=a_i$ for most values of $y$ if $C_{i,j}^{1,q}\neq 0$ and similarly for  $C_{i,j}^{p,1}$ at $y=b_j$.

\subsection{Examples in two dimensions}
All that is needed to evolve a given piecewise polynomial $u(x,y)$ is the location and amount of the discontinuities in  $u$ or any of its derivatives  $u^{p,q}(x,y)$. This information is contained in the list of junction-points of discontinuity $(a_i,b_j)$ and the change $C_{i,j}^{p,q}$ due to jumps in $u^{p,q}(x,y)$ at $(a_i,b_j)$.

\subsubsection{Initial long rectangular distribution:} For an initial distribution with density\,=\,1 over the interior of a rectangle of length $10a$ and width $2a$ the evolution is simple: \mbox{$[\chi(x+a,t)-\chi(x-a,t)][(\chi(y+5a,t)-\chi(y-5a,t)]$}.

\begin{figure}[h!]
\begin{centering}
\includegraphics[width=0.9\textwidth]{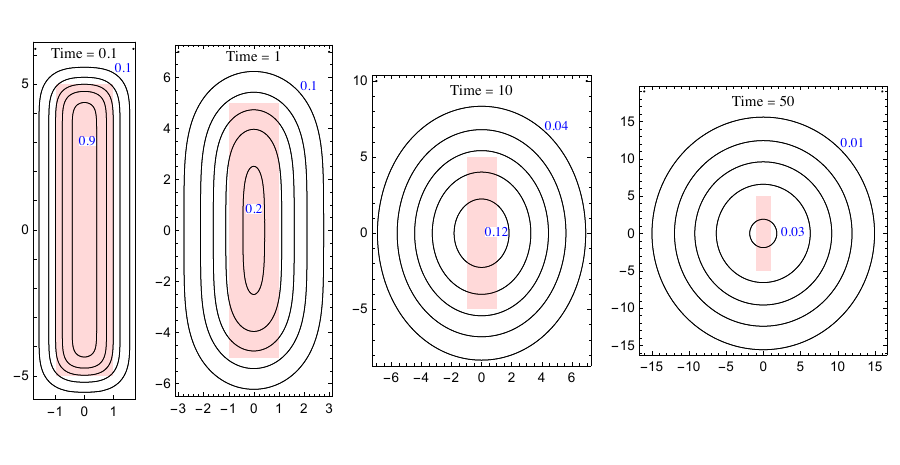} 
 \caption[left]{Contours of the evolution of an initial constant distribution over a long rectangle, shown at four times. The length of the rectangle is $10a$ and the width is $2a$ where $a$ is the unit of length, while the unit of time is $a^2/\kappa$. At each time, the initial rectangle is shown shaded (pink) in the background. The initial density in the rectangle is taken to be 1 and the numbers adjacent to the first and last of the equally-spaced contours give the value of the density at that contour.}
\end{centering}  
\label{fig:3}
\end{figure}

\subsubsection{Bilinear interpolation of initial data on a rectangular mesh:}
The simplest interpolation of a 2D data set is bilinear; the four data points at the corners of each rectangle define a bilinear polynomial $u(x,y)$, that is a linear combination of $x, y, xy$ and a constant. This results in a ruled surface connecting the four data points. Applying this to the whole data set produces a surface that is continuous but with a probable change of slope when crossing each mesh line; but the continuity of the interpolation means that all $C_{i,j}^{1,0}$ and $C_{i,j}^{0,1}$ must be zero. The only non-zero value is $C_{i,j}^{1,1}$ and this will normally be different for each rectangle.

For the rectangle \{i,j\}, solving for the coefficient $c_{i,j}$ of $xy$ results in 
\begin{equation}
c_{i,j} = (u_{i-1,j-1}-u_{i,j-1}+u_{i,j}-u_{i-1,j})/(a_i-a_{i-1})(b_j-b_{j-1}).
\end{equation}
Then, at each junction $(i,j)$, the value of $C_{i,j}^{1,1}$ can be calculated by alternately adding and subtracting $c$ for the four surrounding rectangles:
 \begin{equation}
C^{1,1}_{i,j} = (c_{i-1,j-1}-c_{i,j-1}+c_{i,j}-c_{i-1,j}).
\end{equation}
An equivalent method is to use the data values directly: for equal spacing of the mesh, at any junction the value of $C^{1,1}$ will be four times the data value there, minus twice the sum of the values at the four nearest junctions, plus the sum of the four next-nearest.\cite{A} 

As an example, the function $f(x,y)=1 - \frac{1}{2} (x^2 + y^2 + 2 x)^2 - (x^2 + y^2)$ is used  in the region where $f(x,y)>0$. The contours of this function are shown is Fig.2(a). A rectangular mesh with variable spacing is chosen to reduce the errors in interpolating and fitting to a new boundary on the mesh.
The data was taken to equal the values of $f(x,y)$ at the mesh junctions but zero on the boundary.
Fig.2 also displays the bilinear interpolation and its evolution at time $t=0.001$. The unit of time is $a^2/\kappa$, where $a$ is the unit of length used here for $x$ and $y$.

\begin{figure}[h!]
\begin{centering}
\includegraphics[width=0.92\textwidth]{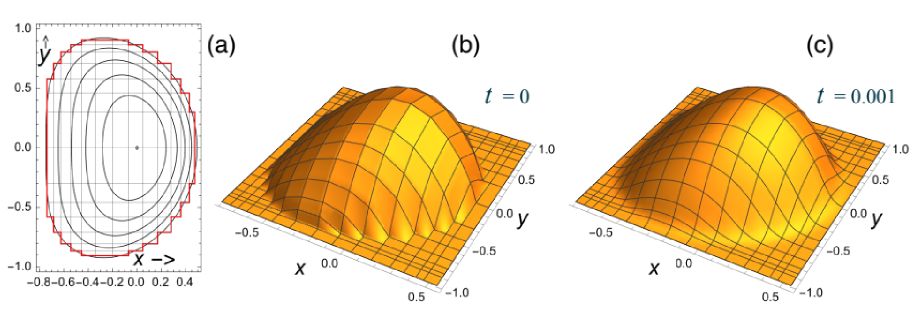} 
 \caption[left]{Evolution of a bilinear interpolation of a 2D data set. (a) Contours (interval 0.2) of $f(x,y)$ and the chosen rectangular mesh. The boundary is shown with heavy (red) lines. (b) The initial bilinear interpolation. (c) The evolution at time $t=0.001$; little has changed, but the corrugations have vanished.}
\end{centering}  
\label{fig4}
\end{figure}

\section{Three-dimensional media}
\vspace{-1mm}
Extending from two to three dimensions is straightforward. For an initial state $\phi(x,y,z)$ the evolution can be found from the integral
\begin{equation}\label{eq:ev3D}
\phi(x,y,z,t)=\int\!\!\!\int\!\!\!\int\!\chi(x-x',t)\,\chi(y-y',t)\,\chi(z-z',t) \phi(x',y',z',0) \,\textrm{d}x'\,\textrm{d}y'\,\textrm{d}z'.
\end{equation}
If the initial state $\phi(x,y,z,0)$ is a single polynomial $u(x,y,z)$ of degree $n$ truncated to a  paralleliped $S=\{a_-\leq x\leq a_+,\;b_-\leq y\leq b_+,\;c_-\leq z\leq c_+\}$, the integral can be converted to a sum using Taylor's theorem:
\begin{eqnarray}
\hspace{-15mm}\sum_{p=0}^n\sum_{q=0}^n\sum_{r=0}^n  f_p(x-a_\alpha)f_q(y-b_\beta)f_r(z-b_\gamma) 
u^{p,q,r}(a_\alpha,b_\beta,c_\gamma) =\nonumber\\ 
\hspace{25mm}\frac{1}{8}\sgn(x-a_\alpha)\sgn(y-b_\beta)\sgn(z-c_\gamma) u(x,y,z),
\end{eqnarray}
for all $x, y, z$, where $u^{p,\,q,\,r}_{\alpha,\beta,\gamma}:=u^{p,q,r}(a_\alpha,b_\beta,c_\gamma)$ and $\alpha, \beta, \gamma$ indicate which corner of $S$ is used as the origin of the Taylor series. Therefore
\vspace{-2mm}
\begin{equation}\label{eqn:ev2Dff}
\sum_{p=0}^n \sum_{q=0}^n  \sum_{r=0}^n\sum_{\alpha=-}^+ \sum_{\beta=-}^+ \sum_{\gamma=-}^+\sigma_{\alpha,\beta,\gamma}
 f_p(x-a_\alpha)f_q(y-b_\beta)f_r(z-c_\gamma)u^{\,p,\,q,\,r}_{\alpha,\beta,\gamma}\!=\!\!
 \begin{cases}u(x,y,z)\,\textrm{in}\, S \\
0\;\textrm{outside}\,\,S,\end{cases}
\end{equation}
where $\sigma_{\alpha,\beta,\gamma}=-\alpha\beta\gamma$, showing that the terms from the $(-,-,-)$-corner (with the lowest values of $x, y,$ and $z$) are added, while those from the three corners nearest to it are subtracted, as are those from the (+,+,+)-corner, while the other three corners are added. The evolution of the truncated polynomial is then
\vspace{-2mm}
\begin{equation}\label{eqn:ev2Dchi}
\phi(x,y,z,t)\!=\!
\sum_{p=0}^n \sum_{q=0}^n \sum_{r=0}^n \sum_{\alpha=-}^+ \sum_{\beta=-}^+
\sum_{\gamma=-}^+\sigma_{\alpha,\beta,\gamma}
\, \chi_p(x-a_\alpha,t)\,\chi_q(y-b_\beta,t)\,\chi_r(z-c_\gamma,t)\,u^{\,p,\,q,\,r}_{\alpha,\beta,\gamma}.
\end{equation}

For $\phi(x,y,z,0)$ defined by a set of polynomials $u_{i,j,k}(x,y,z)$, each within its cuboid $S_{i,j,k}$ with $a_{i-1}<x<a_i, \,b_{j-1}<y<b_j, \,c_{k-1}<z<c_k$,  all within an enclosing region $a_0<x< a_{Nx},\,b_0<y< b_{Ny},\,c_0<z< c_{Nz}$, and with $n$ denoting the largest degree of all the polynomials, the evolution is
\vspace{-2mm}
\begin{equation}\label{eqn:phi_chi3}
\phi(x,y,z,t)=\sum_{i=1}^{Nx} \sum_{j=1}^{Ny} \sum_{k=1}^{Nz}
 \sum_{p=0}^n \sum_{q=0}^n \sum_{r=0}^n
  C_{i,j,k}^{p,q,r}\,\chi_p(x-a_i,t)\,\chi_q(y-b_j,t)\,\chi_r(z-c_k,t),
\end{equation}

\vspace{-4mm}
\noindent where 
\begin{equation}
C_{i,j,k}^{p,q,r}=
\sum_{\alpha=-}^+ \sum_{\beta=-}^+ \sum_{\gamma=-}^+\sigma_{\alpha,\beta,\gamma}\,
u_{\alpha,\,\beta,\,\gamma}^{\;p,\,q,\,r}.
\end{equation}

Continuity of $\phi(x,y,z,0)$ at $(i,j,k)$ implies that $C_{i,j,k}^{p,q,r}=0$ if any of $p,q,r$ are zero. More generally, continuity of the $p,q,r$-derivative of $\phi(x,y,z,0)$ implies that  $C_{i,j,k}^{p,q,r}=0$ for all points $(i,j,k)$ where the continuity applies. In many cases some continuity conditions on derivatives do not apply at the outer borders of the spline.

\pagebreak
\vspace{-10mm}
\subsection{Spherically-symmetric distributions are a 1D problem}
The Laplacian for a spherically-symmetric function $f(r)$ is $\nabla^2f(r)=r^{-1}\partial_r [\partial_r\big(r\,f(r)\big)]$ and therefore the heat equation becomes $\partial_t \big(r\,\phi(r,t)\big) =\kappa\, \partial_{r,r} \big(r\,\phi(r,t)\big)$, that has the same form as the 1D heat equation. If the initial distribution $u(r)$ is within a sphere of radius $a$, the corresponding 1D distribution is the antisymmetric function $v(x,0)=x\,u(|x|)$ for $-a<x<a$ and zero otherwise. Then $\phi(r,t)=r^{-1}v(r,t)$.

With spherical symmetry, the `mass' $M=\int\!\phi(r,t)\,4\pi r^2\,\textrm{d}r$ does not change with time. The Gaussian $G(x)\!=\!M(\pi\,\alpha^2)^{-3/2}\exp(-r^2/\alpha^2)$ has mass $M$ and the radius with half the mass inside is $w_G\!\approx 1.087652\,\alpha$. For an initial state with m-width $w$ the Gaussian with the same m-width has $\alpha=w/w_G$. The exact evolution of  $G(x)$ is
\vspace{-2mm} 
\begin{equation}
G(x,t)=\frac{M}{[\pi(\alpha^2+4\kappa t)]^{3/2}}\exp(-\frac{x^2}{\alpha^2+4\kappa t}).
\end{equation}

\subsubsection{Simple examples of spherically-symmetric distributions.} The simplest example is a sphere of radius $a$ with constant temperature (or density) $d$ over its interior: $u(r)=d$ for $r<a$ and zero otherwise. The equivalent antisymmetric one-dimensional distribution is $v(x)=d\,x$ for $-a<x<a$ and zero otherwise, and this can be expressed as 
$v(x)= d\,[- a\,f_0(x+a)+\,f_1(x+a) - a\,f_0(x-a)-\,f_1(x-a)]$.
 The evolution of $u$ is then $\phi(r,t)=r^{-1}d\,[-a\,\chi_0(x+a,t)+\chi_1(x+a,t) -a\,\chi_0(x-a,t)-\chi_1(x-a,t)]$. The mass is $M=\frac{4}{3}d\,\pi a^3$ and the radius with half the mass inside is $w=a^{-3/2}\approx 0.7937\,a$. The evolution is shown in Fig.5(a).
 
The density $u(r)\!=\!d\,(1-r/a)$ for $r\!<\!a$ and otherwise zero has the one-dimensional equivalent $v(x)=d\,x(1-|x|/a)$. Using $\{a_i\}=(-a,0,a)$ and $\{C_i^2\}=2d\,a^{-1}(1,-2,1)$, $v(x)$ can be expressed as $\sum_{i=1}^3 C^2_i f_2(x-a_i)-f_1(x-a_1)+f_1(x-a_3)$. Then $u(r)=v(r)/r$ and its evolution is
$u(r,t)=d\,r^{-1}[\sum_{i=1}^3C^2_i \,\chi_2(r-a_i)-\chi_1(r-a_1)+\chi_1(r-a_3)]$. The mass $M=\frac{1}{3}d\,\pi a^3$ and the radius with half the mass inside is $w\approx 0.61427243\,a$. The evolution of this initially conical distribution is shown in Fig.5(b).

\vspace{-3mm}
\begin{figure}[h!]
\begin{centering}
\includegraphics[width=0.92\textwidth]{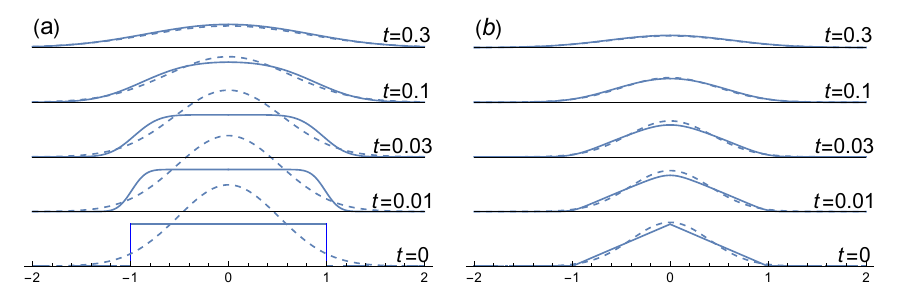} 
\vspace{-3mm}

\caption[left]{The evolution of two spherically-symmetric distributions inside a sphere: (a) a constant density for $r<a$, (b) density $1-r/a$ for $r<a$. The results (solid curves) are shown for a sequence of times, together with the Gaussian (dashed) with the same initial mass and radius $w$ with mean deviation of mass from the centre. The units of length and time are $a$ and $w^2/\kappa$. The mass in (a) is four times that in (b).}
\end{centering}  
\label{fig5}
\end{figure}

\pagebreak
\subsection{A 3D separable tri-quadratic polynomial}
The evolution of a distribution of the form $U(x, y, z)=X(x)Y(y)Z(z)$ is essentially three separate 1D problems: the evolution is simply the product of the three 1D evolutions. Thus, for the initial distribution $U(x, y, z)=(a^2-x^2) (b^2-y^2)(c^2-z^2)$ inside the cuboid $|x|<a,\,|y|<b,\,|z|<c$, the evolution is $\Phi(x, y, z, t)=\phi_a(x, t)\,\phi_b(y, t)\,\phi_c(z, t)$, where
\vspace{-4mm}
\begin{equation}
\phi_a(x, t)=2 a \chi_1(x + a, t) + 2 a \chi_1(x - a, t) - 2 \chi_2(x + a, t) + 
 2 \chi_2(x - a, t).
\end{equation}

\vspace{-2mm}\noindent
[The only discontinuities in $(a^2-x^2)$ are in the first and second derivatives at the edges.]
3D distributions are hard to display, so contours of the cross-section with $z=0$ are shown in Fig.6 with $a=c=1, b=2$. This is similar to the evolution of the 2D distribution $(a^2-x^2) (b^2-y^2)$, but the there is an extra factor arising from $\phi_c(z, t)$ with $z=0$. For two later times, the asymptotic Gaussian $G(x, y, z, t)=M\,G_x(x, t)\,G_y(y, t)\,G_z(z, t)$ is also shown, with
$G_x(x,t)=[\pi(\alpha_x^2+4\kappa t)]^{-1/2}\exp\big(-x^2/(\alpha_x^2+4\kappa t)\big)$. The mass $M=(\frac{4}{3}\,a\,b\,c)^3$ and $\alpha_x \approx a\,w/0.477, \alpha_y \approx b\,w/0.477; \alpha_z \approx c\,w/0.477$, where the m-width of the 1D distribution $a^2-x^2$ is $w \approx 0.6946\,a$.

\vspace{-2mm}
\begin{figure}[h!]
\begin{centering}
\includegraphics[width=0.9\textwidth]{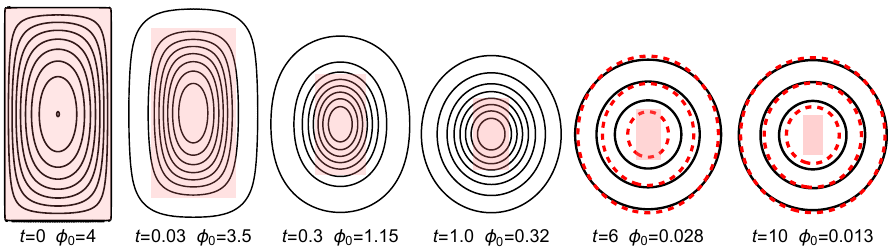} 
\caption[left]{Contours of the evolution of a simple tri-quadratic distribution at various times. The contour interval is one eighth of the maximum value $\phi_0$ at the centre of the cuboid for the first four times and one quarter for the last two times, where the asymptotic Gaussian is also shown as dashed (red). For $t=0.03$ and $t=0.3$ there is an extra outer contour at the value $0.01$. The unit of time is $a^2/\kappa$. The shaded (pink) rectangle in each diagram has the size of the initial distribution.}
\end{centering}  
\label{fig6}
\end{figure}

\subsection{A trilinear piecewise polynomial}
Similar to the 2D case, the only non-zero coefficients $C^{\,p,\,q,\,r}_{i,\,j,\,k}$ have $p=q=r=1$  for a continuous trilinear piecewise polynomial. Then the evolved distribution is
\begin{equation}
\phi(x,\,y,\,z,\,t)=\sum_{i,\,j,\,k}C^{1,\,1,\,1}_{i,\,j,\,k}\,\chi_1(x-a_i,\,t)\,\chi_1(y-b_j,\,t)\,\chi_1(z-c_k,\,t),
\end{equation}
 
 \vspace{-2mm}\noindent
where the sum is taken over all junction points where the $C^{1,\,1,\,1}_{i,\,j,\,k}$ are not zero.
For a mesh with equal spacing, these coefficients can be most easily (but not most efficiently) calculated by taking each junction point $\mathcal{P}$ and summing the weighted values of $u$ at the 27 junctions in its surrounding 8 cubes, with weight 8 at $\mathcal{P}$, -4 for the 6 nearest neighbours, +2 for the 12 next nearest, and -1 for the 8 furthest points \cite{A}.

\vspace{4mm}
Our example has a set of seven samples with mass 4 at the origin, and six other smaller masses each at unit distance from the origin and each on a rectangular axis:
$u_{0, 0, 0} = 4, \;\;u_{1, 0, 0} = 2, \;\;u_{-1, 0, 0} = 2, \;\;u_{0, 0, 1} = 1, 
\;\;u_{0, 0, -1} = 3, \;\;u_{0, 1, 0} = 1, \;\;u_{0, -1, 0} = 2,
$\\
and all other $u_{i, j, k}$ are taken to be zero. We have made a trilinear fit to these data points together with their 74 neighbouring junction points on a rectangular 3D grid with unit spacing. This results in a total of 81 junction points. [This consists of $3^3$ points forming a $3\times3$ cube including the 7 data points, plus $6\times 3^2$ ($3^2$ on each face of the cube) required to keep the boundary one unit away from the sample points.] The initial distribution is non-zero in some immediate neighbourhood of each of the 74 boundary points, but is zero on each such point. Contours of the evolution on the cross-section with $z=0$ are shown in Fig.7.

\vspace{-2mm}
\begin{figure}[h!]
\begin{centering}
\includegraphics[width=0.9\textwidth]{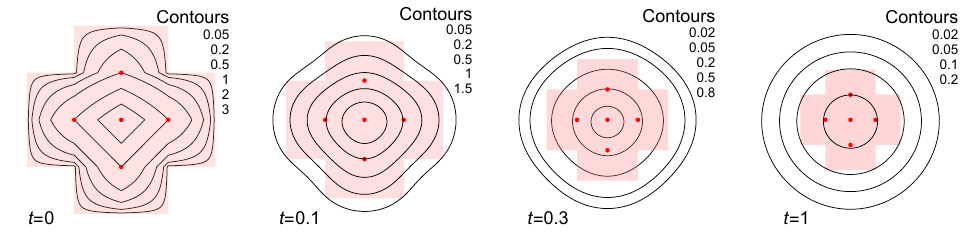} 
\caption[left]{The evolution of a piecewise trilinear polynomial interpolation of seven non-zero data points  in 3D. Contours of the cross-section with $z=0$ is shown with the evolution at three later times.   The data points appear as five (red) points, and the region where the initial interpolation was non-zero is shaded (pink).}
\end{centering}  
\label{fig7}
\end{figure}

\section{Schr\"{o}dinger's equation}
The equation $\partial_t \psi=(i\hbar/2m)\nabla^2 \psi$ for the evolution of a force-free non-relativistic particle, has the same form as the heat equation (\ref{eqn:Heat}) except that the diffusion constant $\kappa$ is imaginary. The quantum propagator

\vspace{-4mm}
\begin{equation}\label{eqn:chiq}
\hspace{1cm}\chi(x,t)=\sqrt{\frac{m}{2\pi i \hbar t}}\exp\big(\frac{\,i m x^2}{2\hbar t}\big)
\end{equation}
is also Gaussian, but is oscillatory and the evolution is dominated by interference. The integral of the propagator is $\chi_0(x,t)=\int_0^x \chi(x',t)\,$d$x'=(2i)^{-1/2}\mathcal{E}(\sqrt{m/\pi \hbar t}x)$, where $\mathcal{E}(z)\!=\!\int_0^z \!\exp(\frac{1}{2}i\pi u^2)\,$d$u\!=\!C(z)+i S(z)$, a complex combination of Fresnel integrals. The repeated integrals of the propagator satisfy a recurrence relation similar to (\ref{eqn:Eseq}) and the limiting form $f_p(x)$ of the evolution functions $\chi_p(x,t)$ as $t\to 0$ is exactly the same as for the Heat equation. The preparation of the evolvable form of a piecewise polynomial wavefunction is exactly the same, as in (\ref{eqn:phi_chi1}, \ref{eqn:phi_chi2}, \ref{eqn:phi_chi3}). The coefficients $C$ and the evolution functions $\chi_p$ may be complex in the quantum case, and the asymptotic form is very different because of interference.\cite{A}.

\section{Discussion}

The procedure described here provides a wide variety of exact evolutions of the heat equation. The standard numerical methods for calculating evolutions from observational data frequently start with a piecewise polynomial fit to the data and this process could be tailored to to a rectangular mesh. Bilinear or trilinear interpolations are simple and efficient; in the context of the heat equation, if the mass distribution closely matches the data, the results of a linear fit can be quite accurate except for the fine detail at very early times. If more accuracy in these details is required, there are good methods to make quadratic fits that are easily evolved. Cubic methods are highly developed but require considerably more calculation, and basis-splines require special handling at the edges to be used with this exact method.

This method of calculating the evolution of a given initial distribution is not subject to accumulating error, as in iterative methods. Furthermore, exact solutions have value in developing understanding of the general behaviour of the evolution and are also useful for testing numerical methods.


\begin{thebibliography}{99}

\bibitem{F} The thermal-conductance equation. Encyclopedia of Mathematics,\\encyclopediaofmath.org.

\bibitem{SW} The Stone-Weierstrass theorem. Encyclopedia of Mathematics,\\encyclopediaofmath.org

\bibitem{M} Abramowitz, M. and Stegun I.A. (eds.): Handbook of Mathematical Functions.\\
 National Bureau of Standards, Washington DC (1964). 7.2.3

\bibitem{V} V\'{a}zquez, J.L.: Asymptotic behaviour methods for the Heat Equation.\\
 arXiv:1706.10034v3, doi.org/10.48550/arXiv.1706.10034 (2017)

\bibitem{K} Kress, R.: Numerical Analysis. Springer, NY (1998). Section 8.3

\bibitem{A} Andrews, M.: The evolution of piecewise polynomial wave functions.\\
Eur. Phys. J. Plus 132(1), (2017), Article 1 of Issue 1, pp14\\
doi 10.11.1140/epjp/i2017-11280-8

\end{thebibliography}
\end{document}